\documentclass[manuscript]{aastex6}
\usepackage{epstopdf} 
\usepackage{epsfig}

\newcommand{\kms}{\ensuremath{\mathrm{km\,s^{-1}}}}
\newcommand{\cms}{\ensuremath{\mathrm{cm\,s^{-1}}}}
\newcommand{\ms}{\ensuremath{\mathrm{m\,s^{-1}}}}

\newcommand{\gcmc}{\ensuremath{\mathrm{g\,cm^{-3}}}}

\newcommand{\micro}{\ensuremath{\mathrm{\mu}}}
\newcommand{\hertz}{\ensuremath{\mathrm{Hz}}}

\newcommand{\nm}{\mathrm{nm}}
\newcommand{\days}{\mathrm{days}}

\newcommand{\gyr}{\mathrm{Gyr}}

\newcommand{\msini}{\ensuremath{m \sin i}}

\newcommand{\teff}{\ensuremath{T_{\mathrm{eff}}}}

\newcommand{\logg}{\ensuremath{\log{g}}}
\newcommand{\vsini}{\ensuremath{v \sin{i}}}
\newcommand{\feh}{\ensuremath{\mathrm{[Fe/H]}}}
\newcommand{\mh}{\ensuremath{\mathrm{[m/H]}}}

\newcommand{\kelvin}{\mathrm{K}}

\newcommand{\rsun}{\ensuremath{R_\sun}}
\newcommand{\msun}{\ensuremath{M_\sun}}

\newcommand{\rstar}{\ensuremath{R_\star}}
\newcommand{\mstar}{\ensuremath{M_\star}}

\newcommand{\rearth}{\ensuremath{R_\earth}}
\newcommand{\mearth}{\ensuremath{M_\earth}}

\newcommand{\rpl}{\ensuremath{R_{p}}}
\newcommand{\mpl}{\ensuremath{M_{p}}}

\newcommand{\rhopl}{\ensuremath{\rho_{p}}}

\newcommand{\kep}{Kepler-20}

\newcommand{\knHIRES}{\ensuremath{30}}
\newcommand{\knHARPSN}{\ensuremath{104}}
\newcommand{\knHARPSNtotal}{\ensuremath{125}}
\newcommand{\knTotal}{\ensuremath{134}}

\newcommand{\kkepMag}{\ensuremath{12.498}}
\newcommand{\kVMag}{\ensuremath{12.51}}
\newcommand{\kra}{\ensuremath{\mathrm{19^h10^m47.^s52}}}
\newcommand{\kdec}{\ensuremath{\mathrm{+42\arcdeg20\arcmin19\farcs4}}}

\newcommand{\ksnr}{\ensuremath{30}}

\newcommand{\kteff}{\ensuremath{5495\pm 50}}
\newcommand{\klogg}{\ensuremath{4.50 \pm 0.10}}
\newcommand{\kloggastero}{\ensuremath{4.446 \pm 0.01}}
\newcommand{\kmh}{\ensuremath{0.07 \pm 0.08}}
\newcommand{\kvsini}{\ensuremath{< 2}}

\newcommand{\kepmstar}{\ensuremath{0.948\pm0.051}}
\newcommand{\keprstar}{\ensuremath{0.964\pm0.018}}
\newcommand{\kage}{\ensuremath{7.6\pm3.7}}

\newcommand{\kjitterHARPS}{\ensuremath{3.93}^{+0.52}_{-0.46}}
\newcommand{\kjitterHIRES}{\ensuremath{3.23}^{+0.68}_{-0.90}}

\newcommand{\krverrintHARPS}{\ensuremath{3.66}}
\newcommand{\krverrintHIRES}{\ensuremath{3.04}}

\newcommand{\krverrtotHARPS}{\ensuremath{5.39}}
\newcommand{\krverrtotHIRES}{\ensuremath{4.45}}

\newcommand{\KBperiod}{\ensuremath{3.69611525_{-0.00000087}^{+0.00000115}}}
\newcommand{\KBtc}{\ensuremath{967.502014_{-0.000217}^{+0.000253}}}
\newcommand{\KBrprstar}{\ensuremath{0.01774_{-0.00003}^{+0.00053}}}
\newcommand{\KBarstar}{\ensuremath{10.34_{-0.32}^{+0.20}}}

\newcommand{\KBinc}{\ensuremath{87.355_{-1.594}^{+0.215}}}
\newcommand{\KBqone}{\ensuremath{0.427_{-0.051}^{+0.120}}}
\newcommand{\KBqtwo}{\ensuremath{0.295_{-0.078}^{+0.134}}}
\newcommand{\KBecc}{\ensuremath{0.03^{+0.09}_{-0.03}}}

\newcommand{\KBsqrtesinw}{\ensuremath{-0.13^{+0.26}_{-0.27}}}
\newcommand{\KBsqrtecosw}{\ensuremath{0.01^{+0.12}_{-0.22}}}

\newcommand{\KBK}{\ensuremath{4.20^{+0.55}_{-0.65}}}
\newcommand{\KBmp}{\ensuremath{9.70^{+1.41}_{-1.44}}}
\newcommand{\KBmpprecision}{\ensuremath{14.7\%}}
\newcommand{\KBrp}{\ensuremath{1.868_{-0.034}^{+0.066}}}
\newcommand{\KBrperr}{\ensuremath{2.7\%}}
\newcommand{\KBrhopl}{\ensuremath{8.2^{+1.5}_{-1.3}}}
\newcommand{\KBa}{\ensuremath{0.0463^{+0.0009}_{-0.0015}}}
\newcommand{\KBteq}{\ensuremath{1105\pm37}}

\newcommand{\KCperiod}{\ensuremath{10.85409089_{-0.00000260}^{+0.00000303}}}
\newcommand{\KCtc}{\ensuremath{971.607955_{-0.000202}^{+0.000248}}}
\newcommand{\KCrprstar}{\ensuremath{0.02895_{-0.00006}^{+0.00029}}}
\newcommand{\KCarstar}{\ensuremath{21.17_{-0.51}^{+0.59}}}

\newcommand{\KCinc}{\ensuremath{89.815_{-0.632}^{+0.036}}}
\newcommand{\KCqone}{\ensuremath{0.393_{-0.038}^{+0.060}}}
\newcommand{\KCqtwo}{\ensuremath{0.408_{-0.069}^{+0.052}}}
\newcommand{\KCecc}{\ensuremath{0.16^{+0.01}_{-0.09}}}
\newcommand{\KCK}{\ensuremath{3.84^{+0.67}_{-0.63}}}
\newcommand{\KCmp}{\ensuremath{12.75^{+2.17}_{-2.24}}}
\newcommand{\KCmpprecision}{\ensuremath{17.3\%}}
\newcommand{\KCrp}{\ensuremath{3.047_{-0.056}^{+0.064}}}
\newcommand{\KCrperr}{\ensuremath{2.0\%}}
\newcommand{\KCrhopl}{\ensuremath{2.5^{+0.5}_{-0.5}}}
\newcommand{\KCa}{\ensuremath{0.0949^{+0.0027}_{-0.0023}}}
\newcommand{\KCteq}{\ensuremath{772\pm26}}

\newcommand{\KCsqrtesinw}{\ensuremath{0.29^{+0.10}_{-0.26}}}
\newcommand{\KCsqrtecosw}{\ensuremath{0.26^{+0.11}_{-0.22}}}

\newcommand{\KDperiod}{\ensuremath{77.61130017_{-0.00011588}^{+0.00012305}}}
\newcommand{\KDtc}{\ensuremath{997.730300_{-0.001577}^{+0.001179}}}
\newcommand{\KDrprstar}{\ensuremath{0.02607_{-0.00021}^{+0.00050}}}
\newcommand{\KDarstar}{\ensuremath{78.23_{-2.25}^{+1.80}}}

\newcommand{\KDinc}{\ensuremath{89.708_{-0.053}^{+0.165}}}
\newcommand{\KDqone}{\ensuremath{0.377_{-0.084}^{+0.188}}}	
\newcommand{\KDqtwo}{\ensuremath{0.205_{-0.083}^{+0.233}}}
\newcommand{\KDecc}{\nodata}
\newcommand{\KDK}{\ensuremath{1.57^{+0.62}_{-0.57}}}
\newcommand{\KDmp}{\ensuremath{10.07^{+3.97}_{-3.70}}}
\newcommand{\KDmpprecision}{\ensuremath{38.1\%}}
\newcommand{\KDrp}{\ensuremath{2.744_{-0.055}^{+0.073}}}
\newcommand{\KDrperr}{\ensuremath{2.3\%}}
\newcommand{\KDrhopl}{\ensuremath{2.7^{+1.1}_{-1.0}}}
\newcommand{\KDa}{\ensuremath{0.3506^{+0.0081}_{-0.0101}}}
\newcommand{\KDteq}{\ensuremath{401\pm13}}

\newcommand{\KDsqrtesinw}{\nodata}
\newcommand{\KDsqrtecosw}{\nodata}

\newcommand{\KGperiod}{\ensuremath{34.940^{+0.038}_{-0.035}}}
\newcommand{\KGtc}{\ensuremath{967.50027^{+0.00058}_{-0.00068}}}
\newcommand{\KGrprstar}{\nodata}
\newcommand{\KGarstar}{\nodata}

\newcommand{\KGinc}{\nodata}
\newcommand{\KGqone}{\nodata}
\newcommand{\KGqtwo}{\nodata}
\newcommand{\KGecc}{\ensuremath{0.15^{+0.01}_{-0.10}}}
\newcommand{\KGK}{\ensuremath{4.10^{+0.61}_{-0.72}}}
\newcommand{\KGmp}{\ensuremath{19.96^{+3.08}_{-3.61}}}
\newcommand{\KGmpprecision}{\ensuremath{16.7\%}}
\newcommand{\KGrp}{\nodata}
\newcommand{\KGrperr}{\nodata}
\newcommand{\KGrhopl}{\nodata}
\newcommand{\KGa}{\ensuremath{0.2055^{+0.0022}_{-0.0021}}}
\newcommand{\KGteq}{\ensuremath{524\pm12}}

\newcommand{\KGsqrtesinw}{\ensuremath{0.15^{+0.12}_{-0.31}}}
\newcommand{\KGsqrtecosw}{\ensuremath{0.23^{+0.11}_{-0.29}}}

\shorttitle{Kepler-20}
\shortauthors{Buchhave et al.}

\begin{document}

\title{A 1.9 Earth radius rocky planet and the discovery of a non-transiting planet in the Kepler-20 system\footnote{Based on observations made with the Italian Telescopio Nazionale Galileo (TNG) operated on the island of La Palma by the Fundaci\'on Galileo Galilei of the INAF (Istituto Nazionale di Astrof\'isica) at the Spanish Observatorio del Roque de los Muchachos of the Instituto de Astrof\'isica de Canarias.}}

\author{
Lars~A.~Buchhave\altaffilmark{1}
Courtney D. Dressing\altaffilmark{2},
Xavier Dumusque\altaffilmark{3},
Ken Rice\altaffilmark{4}, 
Andrew Vanderburg\altaffilmark{5}, 
Annelies Mortier\altaffilmark{6}, 
Mercedes Lopez-Morales\altaffilmark{5},
Eric Lopez\altaffilmark{4},
Mia S. Lundkvist\altaffilmark{7,8}, 
Hans Kjeldsen\altaffilmark{7}, 
Laura Affer\altaffilmark{9},
Aldo S. Bonomo\altaffilmark{10},
David Charbonneau\altaffilmark{5},
Andrew Collier Cameron\altaffilmark{6}, 
Rosario Cosentino\altaffilmark{11}, 
Pedro Figueira\altaffilmark{12}, 
Aldo F. M. Fiorenzano\altaffilmark{11}, 
Avet Harutyunyan\altaffilmark{11}, 
Rapha\"elle D. Haywood\altaffilmark{5}, 
John Asher Johnson\altaffilmark{5},
David W. Latham\altaffilmark{5},
Christophe Lovis\altaffilmark{3},  
Luca Malavolta\altaffilmark{13,14},
Michel Mayor\altaffilmark{3}, 
Giusi Micela\altaffilmark{9}, 
Emilio Molinari\altaffilmark{11,15},
Fatemeh Motalebi\altaffilmark{3},
Valerio Nascimbeni\altaffilmark{13}, 
Francesco Pepe\altaffilmark{3},
David F. Phillips\altaffilmark{5}, 
Giampaolo Piotto\altaffilmark{13,14}, 
Don Pollacco\altaffilmark{16}, 
Didier Queloz\altaffilmark{3,17}, 
Dimitar Sasselov\altaffilmark{5},  
Damien S\'egransan\altaffilmark{3},
Alessandro Sozzetti\altaffilmark{10}, 
St\'ephane Udry\altaffilmark{3},
Chris Watson\altaffilmark{18}
}

\altaffiltext{1}{Centre for Star and Planet Formation, Natural History Museum of Denmark \& Niels Bohr Institute, University of Copenhagen, \O ster Voldgade 5-7, DK-1350 Copenhagen K, Denmark; \href{mailto:buchhave@nbi.ku.dk}{buchhave@nbi.ku.dk}}
\altaffiltext{2}{NASA Sagan Fellow, Division of Geological and Planetary Sciences, California Institute of Technology, Pasadena, CA 91125, USA}
\altaffiltext{3}{Observatoire Astronomique de l'Universit\'e de Gen\`eve, 51 ch. des Maillettes, 1290 Versoix, Switzerland}
\altaffiltext{4}{SUPA, Institute for Astronomy, University of Edinburgh, Royal Observatory, Blackford Hill, Edinburgh, EH93HJ, UK}
\altaffiltext{5}{Harvard-Smithsonian Center for Astrophysics, 60 Garden Street, Cambridge, Massachusetts 02138, USA}
\altaffiltext{6}{SUPA, School of Physics \& Astronomy, University of St. Andrews, North Haugh, St. Andrews Fife, KY16 9SS, UK}
\altaffiltext{7}{Stellar Astrophysics Centre, Department of Physics and Astronomy, Aarhus University, Ny Munkegade 120, 8000, Aarhus C, Denmark}
\altaffiltext{8}{Zentrum für Astronomie der Universität Heidelberg, Landessternwarte, Königstuhl 12, 69117 Heidelberg, Germany}
\altaffiltext{9}{INAF - Osservatorio Astronomico di Palermo, Piazza del Parlamento 1, 90124 Palermo, Italy}
\altaffiltext{10}{INAF - Osservatorio Astrofisico di Torino, via Osservatorio 20, 10025 Pino Torinese, Italy}
\altaffiltext{11}{INAF - Fundaci\'on Galileo Galilei, Rambla Jos\'e Ana Fernandez P\'erez 7, 38712 Bre\~na Baja, Spain}
\altaffiltext{12}{Instituto de Astrof\' isica e Ci\^encias do Espa\c{c}o, Universidade do Porto, CAUP, Rua das Estrelas, PT4150-762 Porto, Portugal}
\altaffiltext{13}{Dipartimento di Fisica e Astronomia ``Galileo Galilei", Universita'di Padova, Vicolo dell'Osservatorio 3, 35122 Padova, Italy}
\altaffiltext{14}{INAF - Osservatorio Astronomico di Padova, Vicolo dell'Osservatorio 5, 35122 Padova, Italy}
\altaffiltext{15}{INAF - IASF Milano, via Bassini 15, 20133, Milano, Italy}
\altaffiltext{16}{Department of Physics, University of Warwick, Gibbet Hill Road, Coventry CV4 7AL, UK}
\altaffiltext{17}{Cavendish Laboratory, J J Thomson Avenue, Cambridge CB3 0HE, UK}
\altaffiltext{18}{Astrophysics Research Centre, School of Mathematics and Physics, Queen’s University Belfast, Belfast BT7 1NN, UK}

\begin{abstract}

Kepler-20 is a solar-type star ($V = 12.5$) hosting a compact system of five transiting planets, all packed within the orbital distance of Mercury in our own Solar System. A transition from rocky to gaseous planets with a planetary transition radius of $\sim 1.6~\rearth$ has recently been proposed by several publications in the literature \citep{rogers_most_2015,weiss_mass-radius_2014}. Kepler-20b ($\rpl \sim 1.9~\rearth$) has a size beyond this transition radius, however previous mass measurements were not sufficiently precise to allow definite conclusions to be drawn regarding its composition. We present new mass measurements of three of the planets in the Kepler-20 system facilitated by 104 radial velocity measurements from the HARPS-N spectrograph and 30 archival Keck/HIRES observations, as well as an updated photometric analysis of the Kepler data and an asteroseismic analysis of the host star ($\mstar = \kepmstar~\msun$ and $\rstar = \keprstar~\rsun$). Kepler-20b is a $\KBrp~\rearth$ planet in a $3.7~\textrm{day}$ period with a mass of $\KBmp~\mearth$ resulting in a mean density of $\KBrhopl~\gcmc$ indicating a rocky composition with an iron to silicate ratio consistent with that of the Earth. This makes \kep b the most massive planet with a rocky composition found to date. Furthermore, we report the discovery of an additional non-transiting planet with a minimum mass of $\KGmp~\mearth$ and an orbital period of $\sim 34~\days$ in the gap between Kepler-20f ($P \sim 11~\days$) and Kepler-20d ($P \sim 78~\days$).

\end{abstract}

\section{Introduction}
\label{sec:intro}

With the discovery of thousands of small transiting exoplanets by dedicated space-based missions like NASA’s Kepler Mission \citep{batalha_planetary_2013} and ESA’s CoRoT Mission \citep{auvergne_corot_2009}, a new class of small planets has emerged. In contrast to the hot-Jupiters, these small planets ($<4~\rearth$) are astonishingly common in our Galaxy \citep{burke_terrestrial_2015, dressing_occurrence_2013,dressing_occurrence_2015, howard_planet_2012, petigura_prevalence_2013}. We now know that the majority of stars harbor small exoplanets and that nearly two-thirds of the planets discovered by Kepler range in size between 1 and 4~\rearth. This unexpected population of super-Earths and sub-Neptunes - larger than Earth but smaller than Neptune (3.9~\rearth) - is completely absent in our own Solar System, which furthermore lacks planets with orbital periods shorter than the 88 day orbit of Mercury. 

Although we know of thousands of these small transiting exoplanets, only a few of these currently have precise mass measurements (precision better than 20\%) allowing us to distinguish between different compositional models. Precise masses and resulting bulk densities are especially important for small planets, since a wide diversity of planet compositions are possible including rocky terrestrial planets with compact atmospheres and rocky cores with significant fractions of volatiles, like water and methane, and/or extended hydrogen/helium envelopes. A transition from rocky to gaseous planets has been proposed to occur at planetary radii of around $1.5-1.7~\rearth$ by a number of authors \citep[e.g.][]{weiss_mass-radius_2014,rogers_most_2015}.

The \kep\ system is particularly interesting because \kep b has a radius ($R_{p,b} = \KBrp~\rearth$) beyond the proposed transition to planets with a significant fraction of volatiles and is amenable to precise determination of its bulk density. Furthermore, the \kep\ systems is intriguing in terms of its formation history, because of its compact nature and the size of the planets alternating between smaller and larger planets in the interior of the system.

Four of the transiting planets in the \kep\ system were first announced as candidate planets in \cite{borucki_characteristics_2011-1}. \cite{gautier_kepler-20:_2012} subsequently validated three of the planets, including mass measurements of \kep b ($M_{p,b} = 8.7^{+2.1}_{-2.2} \mearth$) and \kep c ($M_{p,c} = 16.1^{+3.3}_{-3.7} \mearth$). Later \cite{fressin_two_2012} validated two additional Earth-size planets (\kep e and f) increasing the total number of planets in the \kep\ system to five, all packed within the orbital distance of Mercury in our own Solar System (see Figure \ref{fig:orbitalconfig} for a graphical presentation of the system architecture).

In this paper, we revisit the mass determination of the planets in the \kep\ system with the goal of improving the masses and radii of the planets in order to allow us to discriminate between different compositional models. In particular, we attempt to resolve whether \kep b has a rocky or gaseous composition, since the mass determination in \citet{gautier_kepler-20:_2012} was not of sufficient precision to draw firm conclusions regarding its composition. We have significantly increased the number of radial velocity (RV) measurements by adding \knHARPSN\ HARPS-N observations to the existing \knHIRES\ HIRES measurements allowing a more precise mass determination. Moreover, the new RV measurements allowed us to discover a non-transiting planet in the system situated in the gap between \kep f and \kep d, which we denote \kep g. Lastly, we performed an updated photometric analysis of the Kepler light curve using all the available data in conjunction with an asteroseismic analysis of the parameters of the host stars, yielding a significantly improved precision on the stellar radius and thus the radii of the planets.

\section{Observations and data reduction}
\label{sec:observations}
We obtained \knHARPSNtotal\ observations of \kep\ (KOI-70, KIC 6850504, 2MASS J19104752+4220194) with the HARPS-N spectrograph on the 3.58~m Telescopio Nazionale Galileo (TNG) located at Roque de Los Muchachos Observatory, La Palma, Spain \citep{cosentino_harps-n:_2012}. HARPS-N is an updated version of the original HARPS spectrograph on the 3.6~m telescope at the European Southern Observatory on La Silla, Chile. HARPS-N is an ultra-stable fiber-fed high-resolution $(R = 115,000)$ spectrograph with an optical wavelength coverage from $383$ to $693~\nm$ designed specifically to provide precise radial velocities. The instrument has significantly improved our understanding of small transiting planets with several precise mass measurements of transiting planets \citep{pepe_earth-sized_2013,dumusque_kepler-10_2014,lopez-morales_rossiter-mclaughlin_2014,bonomo_characterization_2014,dressing_mass_2015,gettel_kepler-454_2016}.
We obtained 61 and 64 observations of \kep\ in the 2014 and 2015 observing seasons, respectively (\knHARPSNtotal\ observations in total). We rejected 21 observations obtained under poor observing conditions where the internal error estimate exceeded $5~\ms$ leaving a total of \knHARPSN\ observations. 

\kep\ has a $m_V = 12.5$ and required 30 minute exposure times to build up an adequate signal to noise ratio (SNR). The average SNR per pixel of the observations at $550~\mathrm{nm}$ is \ksnr\ yielding an average internal uncertainty estimate of $3.66~\ms$. The data were extracted and reduced with the standard HARPS-N pipeline. The extracted spectra were cross-correlated with a delta function mask based on the spectrum of a G2V star \citep{baranne_elodie:_1996,pepe_coralie_2002}. The resulting radial velocities and their $1 \sigma$ errors are shows in Table \ref{tab:rvs} along with their weighted midtime of exposure in BJD UTC, the activity indicators FWHM (full width at half maximum) and $\mathrm{Ca~II}  \log(R'_{\mathrm{HK}})$ with respective uncertainties and SNR at $550~\nm$.

\cite{gautier_kepler-20:_2012} obtained 30 spectra of \kep\ between August 2009 and June 2011 using the HIRES spectrometer on the Keck I 10 meter telescope \citep{vogt_hires:_1994}. With a typical exposure time of 30 to 45 minutes on a telescope with a light collecting power almost a factor of 8 greater than that of the TNG, the average SNR per pixel is 120 yielding an average internal uncertainty of $3.04~\ms$. If the internal errors are taken at face value, the increase in SNR by $400\%$ leads to a increase in RV precision of only $20\%$, when compared to the HARPS-N RVs. In Section \ref{sec:rvanalysis}, we calculate the average internal uncertainty including an added jitter term for each instrument. The conclusion is that the average internal uncertainty of the HIRES RVs is $23\%$ more precise than the HARPS-N RVs, despite the $400\%$ higher SNR.

\begin{deluxetable*}{cccccccc} 
	\tabletypesize{\footnotesize} 
	\tablewidth{16.5cm} 
	\tablecaption{HARPS-N RV measurements. \label{tab:rvs}}
	\tablecolumns{8}
	\tiny
	\tablehead{
		\colhead{BJD - 2,400,000 } &
		\colhead{RV} &
		\colhead{RV error} &
		\colhead{FWHM} &
		\colhead{FWHM error} &
		\colhead{log(R'$_{HK}$)} &
		\colhead{log(R'$_{HK}$) error} &
		\colhead{SNR} \\	
		\colhead{(days)} &
		\colhead{(\ms)} &
		\colhead{(\ms)} &
		\colhead{(\ms)} &
		\colhead{(\ms)} &
		\colhead{(dex)} &
		\colhead{(dex)} &
		\colhead{550 nm} 
	}
	\startdata   
56764.716145   &   -20933.75   &   4.52   &   6874.48   &   10.62   &   -4.82   &   0.06   &   25.80 \\
56765.642308   &   -20932.51   &   3.86   &   6891.10   &   9.07   &   -4.87   &   0.06   &   29.60 \\
56769.649982   &   -20917.87   &   3.76   &   6884.07   &   8.84   &   -4.82   &   0.05   &   28.80 \\
56783.585315   &   -20934.24   &   4.03   &   6874.57   &   9.47   &   -4.94   &   0.07   &   29.10 \\
56784.611099   &   -20925.09   &   3.98   &   6873.34   &   9.35   &   -4.79   &   0.05   &   28.70 \\
	\nodata & \nodata & \nodata & \nodata & \nodata & \nodata & \nodata & \nodata \\
	\enddata
	\tablecomments{Table \ref{tab:rvs} is published in its entirety in the electronic edition of the Astrophysical Journal. A portion is shown here for guidance regarding its form and content.}

\end{deluxetable*}

\begin{deluxetable}{lcc}
	\tabletypesize{\scriptsize}
	\tablecaption{System Parameters for \kep\label{tab:stellar}}
	\tablehead{
		\colhead{Parameter}				& 
		\colhead{Value} 				&
		\colhead{Ref.}					\\
	}
	\startdata
	Right Ascension (J2000) & \kra & 1 \\
	Declination (J2000) & \kdec & 1 \\
	Kepler magnitude & \kkepMag & 1 \\
	V magnitude & \kVMag & 2 \\
	\teff (K)  & \kteff & 3 \\
	\logg  & \kloggastero & 4 \\
	\mh & \kmh & 3 \\
	\vsini\ (\kms)  & \kvsini & 3 \\
	\mstar\ (\msun)  & \kepmstar & 4 \\
	\rstar\ (\rsun)  & \keprstar & 4 \\
	Age (Gyr)  & \kage & 4 \\
	Systemic velocity $\gamma~(\ms)$ & $-20926.9^{+0.54}_{-0.64}$ & 5 \\
	\enddata
	\tablenotetext{1}{\cite{gautier_kepler-20:_2012}}
	\tablenotetext{2}{\cite{lasker_second-generation_2008}}
	\tablenotetext{3}{From SPC analysis in Section \ref{sec:spc}}
	\tablenotetext{4}{From asteroseismic analysis in Section \ref{sec:asteroseis}}
	\tablenotetext{5}{From RV analysis in Section \ref{sec:rvanalysis}}

\end{deluxetable}

\section{Stellar properties}
\label{sec:stellarproperties}

\subsection{Spectroscopy}
\label{sec:spc}

\cite{gautier_kepler-20:_2012} obtained stellar parameters for \kep\ from two HIRES template spectra without the iodine cell that were analyzed using SME \citep{valenti_spectroscopy_1996,valenti_spectroscopic_2005} yielding $\teff = 5455 \pm 44~\kelvin$, $\logg = 4.40 \pm 0.10$, $\feh = 0.01 \pm 0.04$ and $\vsini < 2~\kms$. The Stellar Parameter Classification tool \citep[SPC;][]{buchhave_abundance_2012,buchhave_three_2014} was also used to analyze spectra from FIES, McDonald and HIRES, yielding $\teff = 5563 \pm 50~\kelvin$, $\logg = 4.52 \pm 0.10$, $\mh = 0.04 \pm 0.08$ and $\vsini = 1.8 \pm 0.5~\kms$, consistent with the SME values, except for the effective temperature, which is $108~\kelvin$ higher.

We used SPC\citep{buchhave_abundance_2012,buchhave_three_2014} to derive the stellar parameters of the host star from the \knHARPSNtotal\ high-resolution high SNR HARPS-N spectra (average SNR per pixel of \ksnr). Since SPC does not require a very high SNR, we were able to utilize all the spectra for the analysis. We ran SPC with all parameters unconstrained and with the surface gravity constrained to the value determined from asteroseismology ($\logg = \kloggastero$, see Section \ref{sec:asteroseis}). The surface gravity from the unconstrained SPC analysis, $\logg = \klogg$, is in close agreement with the value from asteroseismology. The unconstrained weighted mean of the SPC results from the individual spectra yielded: $\teff = 5532\pm 50~\kelvin$, $\logg = 4.50 \pm 0.10$, $\mh = 0.08 \pm 0.08$ and $\vsini < 2~\kms$. Imposing a prior on the surface gravity from the asteroseismic analysis yielded the final set of stellar parameters adopted in this paper: $\teff = \kteff~\kelvin$, $\mh = \kmh$ and $\vsini \kvsini~\kms$.

We note that the weighted mean of the SPC classifications with a prior on the surface gravity of the multiple HARPS-N spectra is in agreement with the initial SME and SPC analysis in the discovery paper \citep{gautier_kepler-20:_2012} as well as the adopted effective temperature value in the discovery paper. The effective temperature and metallicity values from SPC were used in the asteroseismic analysis presented in the following section.

\begin{figure*}
	\begin{center}
		\includegraphics[width=0.95\textwidth]{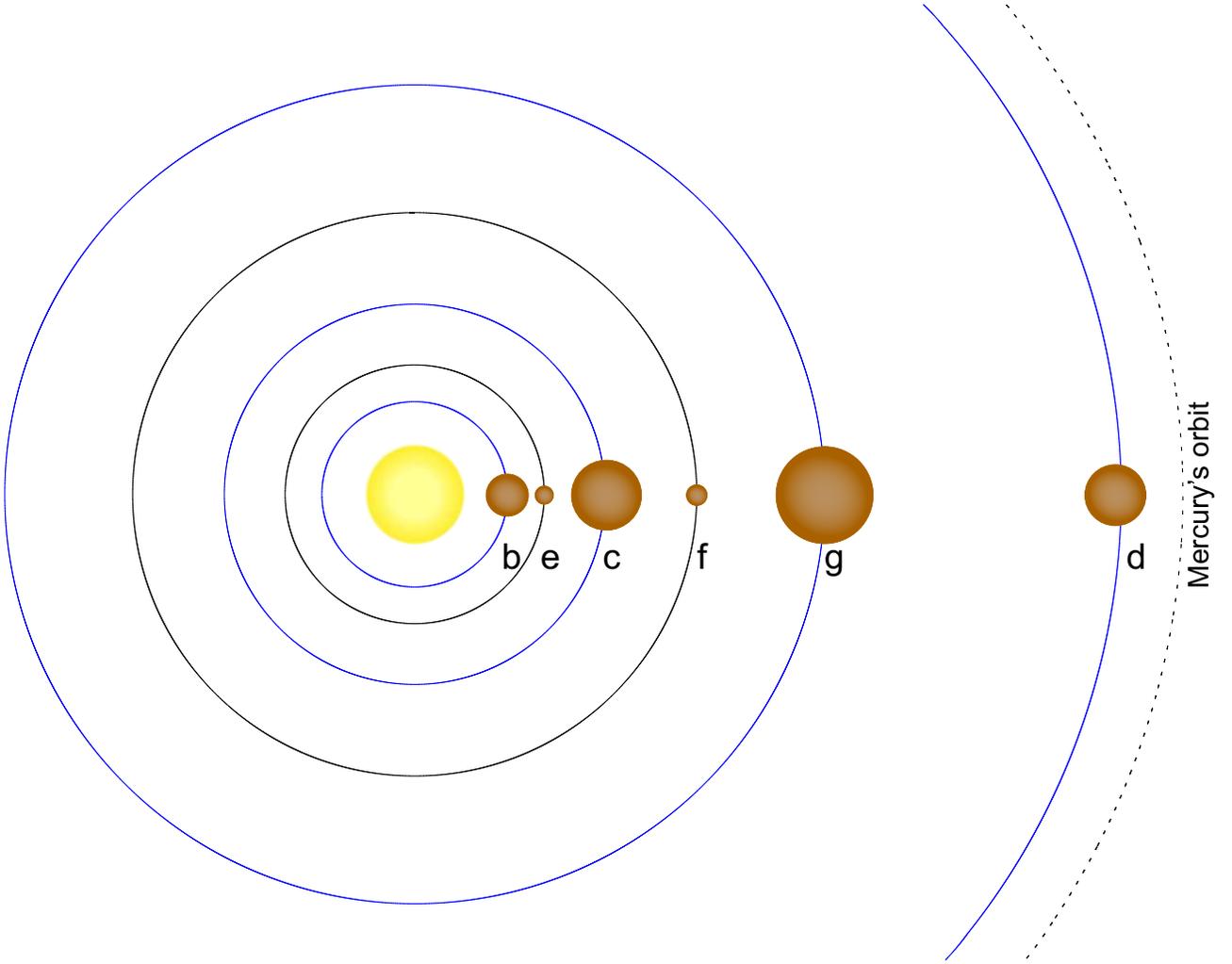}
		\caption{The orbital configuration of the \kep\ system, where all the planets are packed within the orbital distance of Mercury. The orbital distances are to scale and the planet sizes are scaled to the correct size relative to each other, but have been increased significantly (The radius of \kep g has been estimated using its mass and assuming a similar composition as \kep c). The blue lines represent the planets with mass measurements.}
	\end{center}
	\label{fig:orbitalconfig}
\end{figure*}

\subsection{Asteroseismic analysis}
\label{sec:asteroseis}

The power spectrum of Kepler-20 shows low signal-to-noise solar-like oscillations, and as a consequence asteroseismology can be used to infer basic stellar properties \citep[for an introduction to asteroseismology, see for instance][]{aerts_asteroseismology_2010,chaplin_asteroseismology_2013}. Due to the low signal-to-noise ratio of the oscillations, we could not detect the individual oscillation frequencies, but we were able to determine one of the global properties of the oscillations, namely the large frequency separation.

The large frequency separation is the main regularity of the pulsation pattern visible in a power spectrum. This can be seen from the asymptotic relation \citep{tassoul_asymptotic_1980}. The asymptotic relation gives to a good approximation the frequency of a p mode in a star showing solar-like oscillations as a function of the radial order ($n$) and the degree ($\ell$) of the mode:
\begin{equation}
\label{astseis:eq:nu_asymp}
\nu_{n,\ell} = \Delta \nu \left( n + \frac{\ell}{2} + \epsilon \right) - \ell \left( \ell + 1 \right) D_0 \ .
\end{equation}
Here, $D_0$ is a quantity that depends on the sound speed gradient near the core, while $\epsilon$ is a parameter of order unity that is sensitive to the near-surface layers, and $\Delta\nu$ is the large frequency separation. The large separation is the inverse of the sound travel time across the star, and it has been shown to scale with the square root of the stellar mean density; $\Delta\nu \propto \sqrt{\bar{\rho}}$ \citep{ulrich_determination_1986,kjeldsen_amplitudes_1995}.

\begin{figure}[htbp]
	\begin{center}
		\plotone{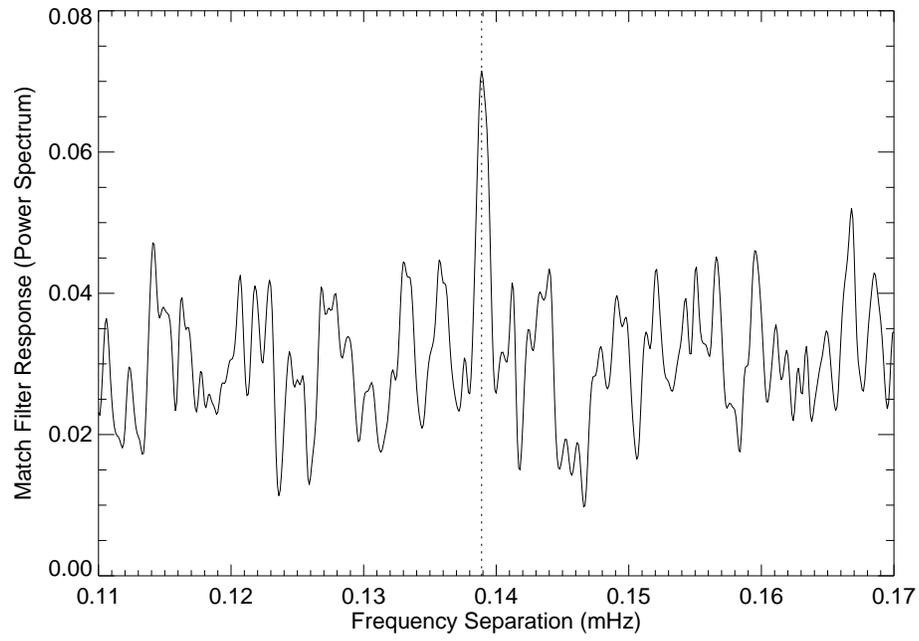}
		\caption{Value of the matched filter response as a function of the large frequency separation. The peak at $138.9 \pm 0.7 \ \micro\hertz$ (marked by the black dotted line) gives the frequency separation found in Kepler-20.}
		\label{fig:MFR_dnu}
	\end{center}
\end{figure}

We determined the large frequency separation of Kepler-20 using the matched filter response function described by \citet{gilliland_asteroseismology_2011}. 
With this technique, the power spectrum is searched for a signal essentially given by the asymptotic relation (see Equation~\ref{astseis:eq:nu_asymp}) for a range of possible values of the input parameters, including $\Delta\nu$. The result is the value of the matched filter response as a function of the frequency separation, which can be seen in Figure \ref{fig:MFR_dnu}. From this it can be found that the large frequency separation of Kepler-20 is $\Delta\nu = 138.9 \pm 0.7 \ \micro\hertz$, which is the frequency separation where the matched filter response takes its largest value. The $1 \sigma$ uncertainty has been determined as the full width at half maximum of the peak in the matched filter response.

In order to derive the basic stellar properties, we combined the determined $\Delta\nu$ with the effective temperature and the metallicity of Kepler-20 and used this as input to the grid-modeling code AME \citep{lundkvist_ame_2014}. Using AME we obtained the stellar parameters given in Table \ref{tab:stellar} (note that AME imposes a lower limit on the uncertainty of the large frequency separation of $1\%$).

It should be noted, that recently NASA Ames released an erratum concerning approximately half of the short cadence targets observed by Kepler\footnote{the erratum can be found at http://keplerscience.arc.nasa.gov/data/documentation/KSCI-19080-002.pdf}, including Kepler-20. The reason for the erratum is that a fundamental error introduced during the calibration of the pixel data was uncovered. The implication for the affected targets is that the data have an increased noise level with the actual increase varying from target to target. The error will be rectified in a future data release, which is consequently expected to improve the signal-to-noise ratio in the power spectrum of Kepler-20. Since the long cadence data for the same targets are not altered by the discovered error, it is possible to get an estimate of the magnitude of the effect by comparing the long cadence and short cadence pixel data. We have done this for representative examples of data from Kepler-20, and we estimate the error to only have a mild impact on the power spectrum of Kepler-20. However, there is little doubt that with the re-processed data, we will be able to obtain a more significant detection of the solar-like oscillations in Kepler-20. The new data will be available in the second half of 2016.

\section{Analysis of the \emph{Kepler} Photometry}
\label{sec:photanalysis}


\citet{gautier_kepler-20:_2012} and \citet{fressin_two_2012} presented properties of the Kepler-20 planets based on analyses of the first two years of \emph{Kepler} photometry. Although Kepler-20 was observed at short cadence (integration time per data point of 58.8~s) beginning in Q3, \citet{gautier_kepler-20:_2012} conducted their analysis using the Q1-Q8 long-cadence photometry (29.426~min integration time) to reduce the computational burden of the analysis. Although fitting the long-cadence photometry is more computationally efficient, the relatively long integration time smears out the shape of planetary transits and increases the degeneracy between parameters in the resulting transit fit. In contrast,  short-cadence photometry better reveals the morphology of transits, particularly the duration and slope of ingress and egress. Consequently, we performed our transit fits using the full set of short-cadence light curves. These light curves were obtained during Q3 - Q17 (18 September 2009 - 7 May 2013) and are divided into 44 separate files with baselines of approximately one month. 

We prepared the data for transit fits by first normalizing each light curve by its median and combining them to form a single time series. We then extracted small segments of the light curve centered around each planetary transit and normalized them using a linear fit to the out-of-transit data. Specifically, we assumed the transit centers and durations reported in the NASA Exoplanet Archive and used least squares minimization to fit a line to the observations acquired $1-3$ transit durations away from the expected transit center. Prior to extracting the transits for each planet, we removed the segments of the light curve containing the transits of the other planets in the system. We rejected transits for which the remaining data coverage was too sparse due to data gaps or the removal of other planetary transits. Our resulting light curve segments contained 308~transits for Kepler-20b, 105~transits for Kepler-20c,  9~transits for Kepler-20d, 188~transits for Kepler-20e, and 58~transits for Kepler-20f.   

Next, we used a Markov chain Monte Carlo (MCMC) analysis to refine the parameters of each planet. Our analysis varied the following parameters: orbital period, $P$; time of transit center, $t_0$; planet-to-star radius ratio, $R_p/R_\star$; semimajor-axis-to-stellar-radius ratio, $a/R_\star$; planetary orbital inclination, $i$; eccentricity, $e$; argument of periastron, $\omega$; and two limb darkening parameters, $q_1$ and $q_2$, which constrained a quadratic limb darkening law in the manner of \citet{kipping_characterizing_2014}. Rather than vary $e$ and $\omega$ directly, we reparameterized $e$ and $\omega$ into two alternative variables. For all planets, we used the ECCSAMPLES package \citep{kipping_bayesian_2014} to transform two uniform variates ($x_e$, $x_\omega$) into ($e$, $\omega$) pairs drawn from a Beta distribution describing the eccentricity distribution of transiting exoplanets \citep{kipping_parametrizing_2013,kipping_bayesian_2014}.

In all cases, we  incorporated uniform priors of \mbox{$0.1 <  P < 200{\, \rm d}$}, \mbox{$100 < t_0 < 200{\, \rm BKJD}$}, \mbox{$0 < R_p/R_\star < 0.1$},  \mbox{$70^\circ < i < 90^\circ$}, \mbox{$0 < q_1 < 1$}, and \mbox{$0 < q_2 < 1$}. We constrained the $a/R_\star$ values by imposing gaussian priors set by the asteroseismic stellar density estimate \citep{seager_unique_2003,sozzetti_improving_2007,torres_improved_2008,ballard_kepler-93b:_2014} and restricted the eccentricities to \mbox{$\leq 0.3$},\mbox{$\leq 0.17$},\mbox{$\leq 0.28$}, \mbox{$\leq 0.28$}, and \mbox{$\leq 0.32$} for planets b, c, d, e, and f, respectively, to comply with the stability constraints described in Section \ref{sec:stability} and prevent overlapping orbits. Relaxing these upper limits does not significantly change the resulting planet radii.

\begin{figure*}[htbp]
	\begin{center}
		\includegraphics[width=0.95\textwidth]{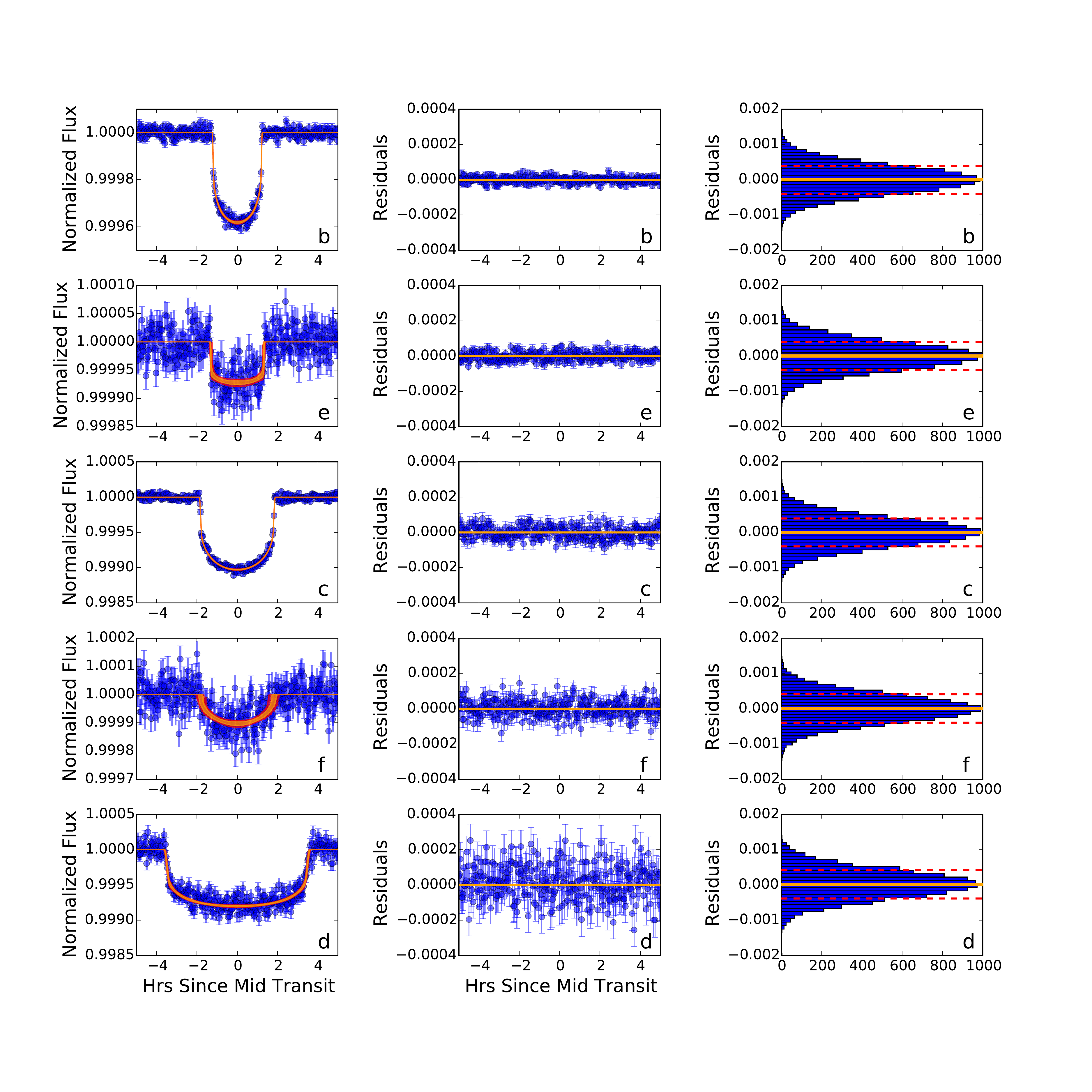}
		\caption{Phase-folded light curves (left panels), phase-folded residuals (center panels), and histograms of residuals (right panels) for fits to transits of Kepler-20b (top), 20e (second from top), 20c (middle), 20f (second from bottom), and 20d (bottom).  In the left and middle panels, the blue data points are short cadence data binned to a 2-minute intervals. The shaded lines in the left panel indicate the $1\sigma$ (red) and $3\sigma$ (orange) confidence flux intervals for our best-fit solutions. In the middle panel, the red line marks zero residuals. In the right panels, the solid orange line mark the median values and the dashed red lines indicate the 16th and 84th percentile of the distributions. The residual histogram was constructed using the individual short cadence data points (not the binned values plotted in the left and middle panels). The vertical scales are different for each planet in the left panel, but standardized for the residuals plots in the middle and right panels.}
	\end{center}
	\label{fig:photfits}
\end{figure*}

We performed the light curve fits in python by employing the {\tt emcee} Affine-Invariant MCMC ensemble sampler package \citep{fortney_framework_2013} to run MCMC and the {\tt BATMAN} package \citep{kreidberg_batman:_2015} to generate light curves using analytic models \citep{mandel_analytic_2002}. We adopted the values provided on the NASA Exoplanet Archive as initial guesses and initialized the MCMC walkers in tight Gaussian distributions surrounding the initial solution. We used $200-500$ walkers per analysis and ran the chains for $13,000 - 45,000$ steps. We discarded the initial $400 - 10000$ steps of each chain as ``burn-in'' and followed the guidance of \citet{fortney_framework_2013} by running the chains for $20-110$ times longer than the autocorrelation time to ensure that the chains were well-mixed. Our best-fit solutions for each planet are provided in Table~\ref{tab:photoparameters} \& \ref{tab:planetparameters} and displayed in Figure~\ref{fig:photfits}; we report the modes of the posterior distributions as the best-fit values and set the errors to encompass 68\% of the values within each distribution. 

In the previous analyses \citep{gautier_kepler-20:_2012,fressin_two_2012}, the reported radii of the planets were $1.91^{+0.12}_{-0.21}\rearth$ (Kepler-20b), $3.07^{+0.20}_{-0.31}\rearth$ (Kepler-20c), $2.75^{+0.17}_{-0.30}\rearth$ (Kepler-20d), $0.868^{+0.074}_{-0.096}\rearth$ (Kepler-20e), and $1.03^{+0.10}_{-0.13}\rearth$ (Kepler-20f). In comparison, our new fits are \mbox{$1.868^{+0.066}_{-0.034}\rearth$}, \mbox{$3.047^{+0.064}_{-0.056}\rearth$}, \mbox{$2.744^{+0.073}_{-0.055}\rearth$}, \mbox{$0.864^{+0.026}_{-0.028}\rearth$}, and \mbox{$1.003^{+0.050}_{-0.089}\rearth$}, respectively. These values are consistent with the previous estimates, but the new errors are smaller by factors of $1.5 - 6.2$.

\subsection{Investigating the Presence of Transit Timing Variations}
Given the likely presence of a sixth, massive and non-transiting planet in the Kepler-20 system, we conducted a search for transit timing variations (TTVs) to test whether this non-transiting planet (or another planet in the system) might be perturbing the transit times of planets  b, c, d, e, and f. We searched for TTVs by generating model light curves based on the best-fit parameters given in Table~\ref{tab:photoparameters}  \& \ref{tab:planetparameters} and comparing each individual transit event to the corresponding model light curve. Holding all other parameters constant, we varied the time of transit over a fine grid of values extending from 90~minutes early until 90~minutes late and recorded the transit center that minimized the chi-squared of the model fit. We then assigned errors on our estimated transit times by determining the range of transit centers that yielded $\Delta \chi^2 < 1$ compared to the best-fit solution. None of the planets displayed coherent TTVs. The non-detection of periodic TTVs in our data is consistent with the earlier null result by \citet{gautier_kepler-20:_2012}.

\section{Analysis of the radial velocity data}
\label{sec:rvanalysis}

In this section, we describe our analysis of the combined RV data from HARPS-N and HIRES totaling \knTotal\ observations gathered from 2009 to 2015. There are five transiting planets in the \kep\ system, however, the two smallest planets, \kep e and \kep f, have sizes below that of the Earth. Even when assuming a rocky composition, which would maximize the resulting RV signal, the estimated semi-amplitudes of these two planets are $18~\cms$ and $19~\cms$, respectively, which is well below the threshold at which we can detect orbital motion. Therefore, we do not include these two planets in our RV analysis.

The estimated age of the system of $\kage~\gyr$ from asteroseismology implies a low stellar activity level. The $\mathrm{Ca~II~HK}$ chromospheric activity index, which is a good indicator for the activity level of solar type stars, yields an average value of $\log(R'_{\mathrm{HK}}) = -4.88 \pm 0.07$. This is similar to the mean activity index of the Sun, which varies between -5.0 and -4.8. Observations of the Sun as a star show an RV RMS of a few meter per second \citep{haywood_sun_2016}, and we do therefore not expect, with the RV precision reached on HARPS-N and HIRES for this faint target, that stellar signals strongly affect our RV measurements. We observe no significant correlation between the radial velocities and the FWHM of the cross correlation function, the bisector span or the $\mathrm{Ca~II~HK}$ index.

The analysis of the photometric data from Kepler in Section \ref{sec:photanalysis} yields very precise constraints on the periods and transit epoch of the transiting planets in the system. It is clear that the period and time of transit are constrained by the photometry itself and the \knTotal\ RV data points do not improve the determination of these parameters. Thus, we fit the RV data separately, using the strong priors on the periods and times of transit from the photometry.

We fitted a model with three Keplerian signals, one for each of the three larger transiting planets (\kep b, \kep c and \kep d), an RV offset for each instrument (HARPS-N and HIRES) and a jitter term for each instrument to account for instrumental systematics not included for in the formal uncertainties and/or stellar induced astrophysical noise. Each planet $j$ is characterized by its semi-amplitude $K_j$, period $P_j$, time of transit $T_{c,j}$, eccentricity $e_j$ and argument of periastron $\omega_j$. The time of transit was shifted to coincide with the average date of the RV observations in order to minimize error propagation from the uncertainty on the periods. Rather than using eccentricity and argument of periastron as free parameters, we used $\sqrt{e_j}\cos{\omega_j}$ and $\sqrt{e_j}\sin{\omega_j}$ as free parameters allowing for a more efficient exploration of parameters space at small eccentricities \citep{ford_improving_2006}.

\begin{figure}
	\begin{center}
		\plotone{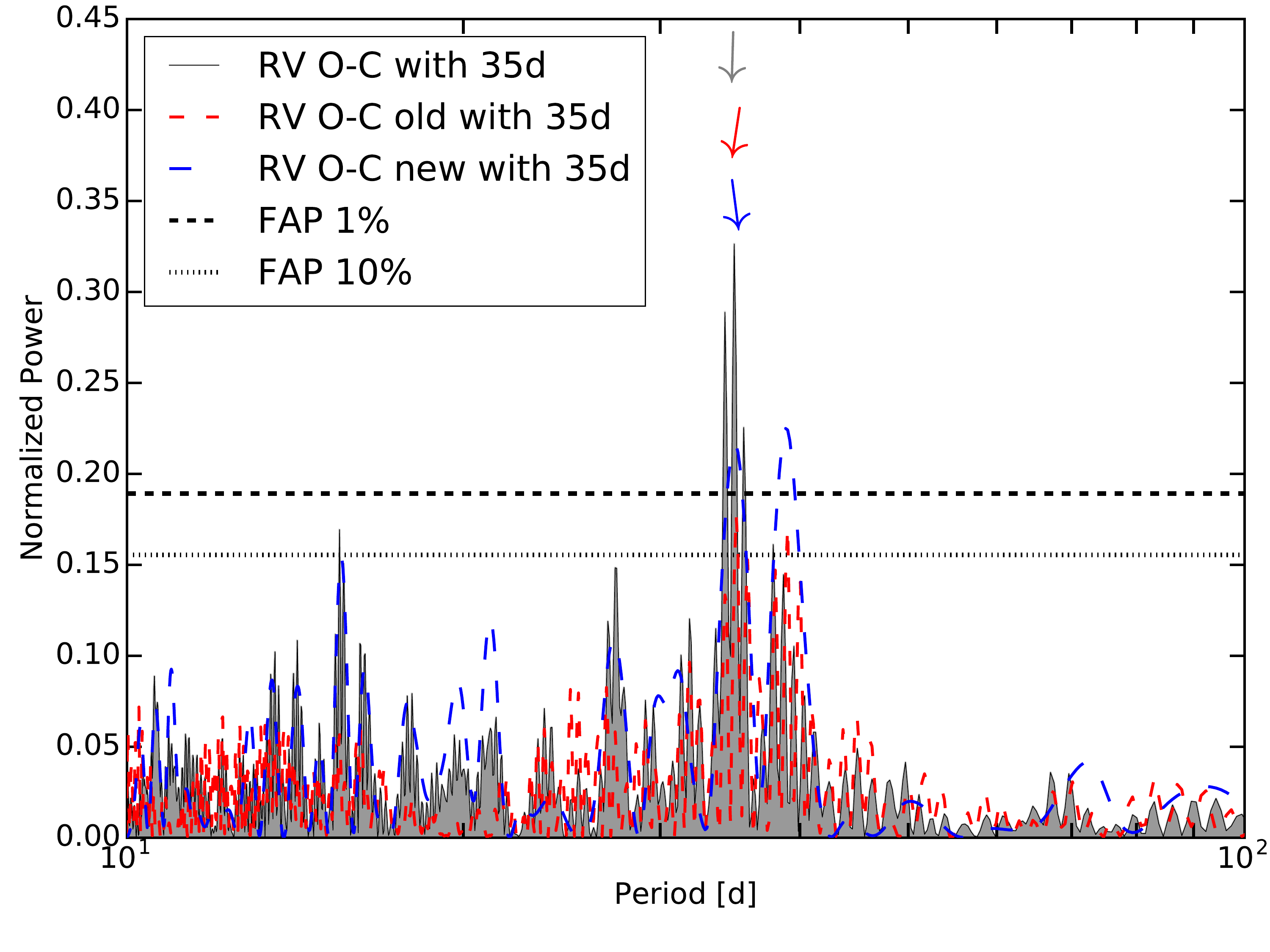}
		\caption{GLS periodogram of the RV residuals (O$-$C) after removing our best fit for Kepler-20b, c, and d. The black shaded periodogram corresponds to the analysis of the entire data set, while the red and blue periodograms correspond to the first and second halves of the data, respectively. The small arrows above the 35-day peak show the phase of the signal, which can rotate by 360 degrees depending on the phase. The phase of the first and second halves of the data are compatible and also compatible with the phase found for the entire data set. Therefore, the 35-day signal retains its phase as a functions of time, which is expected for a planet. The top and bottom horizontal lines correspond to a FAP level of 1\% and 10\%, respectively}
		\label{fig:xdum_fig0}
	\end{center}
\end{figure}

We explored the parameter space using a Bayesian Markov Chain Monte Carlo analysis with a Metropolis-Hastings acceptance criterion. The jitter terms were incorporated into the likelihood function as in \citet{dumusque_kepler-10_2014} and \cite{dressing_mass_2015}. We imposed Gaussian priors on period and time of transit set by the photometric analysis in Section \ref{sec:photanalysis} and used uniform priors on all other free parameters. Furthermore, we required that jitter and semi-amplitude values must be positive and we constrained the eccentricity to be $0 \leq e < 1$.  We performed a ``burn-in'' by running the chain for $10^6$ steps and saved subsequents steps. From the posteriors, we selected the mode of each parameters and assigned uncertainties encompassing $68 \%$ of the posterior closest to the adopted best fit value.
After fitting the three transiting planets, we examined the residual RVs to check for any systematic deviations from the fitted solution. Figure \ref{fig:xdum_fig0} shows a periodogram of the residual RVs (black line). A strong peak is clearly visible at $\sim~35~\days$, indicating the presence of another body in the system or stellar activity caused by spot modulation at the stellar rotational period. As we demonstrate in Section \ref{sec:nontransit}, there is strong evidence that the residual RV signal is caused by a non-transiting planet in the system and not by stellar activity. We therefore denote this planet \kep g and return to the argument for its planetary nature in Section \ref{sec:nontransit}. \kep g has an orbital distance that places its orbit in the gap between \kep f and \kep d (see Figure \ref{fig:orbitalconfig}). To account for this non-transiting planet, we performed a MCMC analysis as before, but added an additional Keplerian signal, three of which have Gaussian priors on period and time of transit and the fourth with uniform priors. The analysis yielded a minimum mass for \kep g of $\sim 21~\mearth$ slightly larger than Neptune in our own Solar System.

The \kep\ system is quite compact with 6 planets all packed within the orbital distance of Mercury and in order for the system to be stable over long timescales, we would naively expect the planets to have rather moderate eccentricities. Since the constraints on the eccentricities of the planets from the RV analysis is weak, we performed a stability analysis of the system using N-body simulations. The result of the analysis, described in detail in Section \ref{sec:stability}, is that the system is indeed stable over long timescales, if the eccentricities of the planets are low. We can use these eccentricity constraints to impose priors on the eccentricities in the RV analysis. We therefore performed a final MCMC analysis of the RVs imposing priors on the eccentricities of planets \kep c, \kep d and \kep g of $e_c \le 0.17$, $e_d \le 0.28$ and $e_g \le 0.16$ (see Section \ref{sec:stability} for details). These constraints are compatible with the eccentricities derived from the RV analysis with uniform priors on the eccentricities, within their uncertainties.

\begin{figure*}
	\begin{center}
		\includegraphics[width=0.95\textwidth]{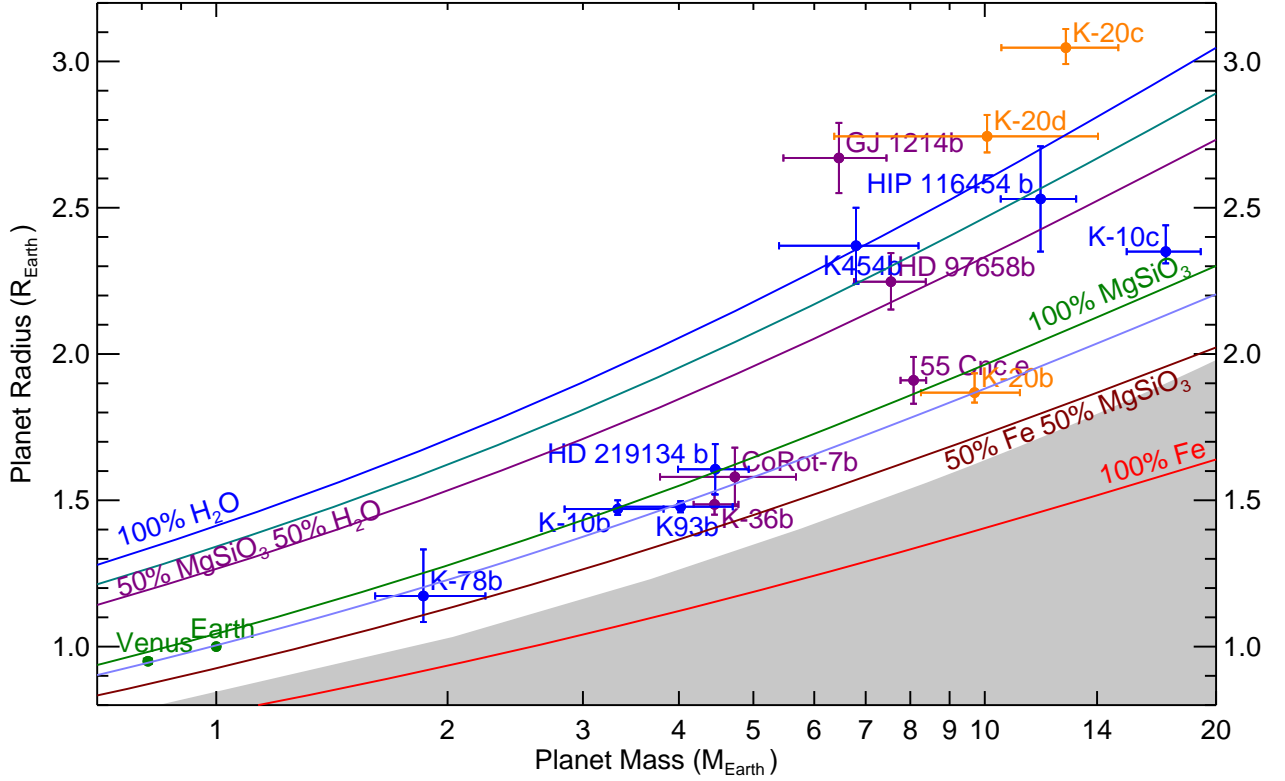}
		\caption{Mass-radius diagram for planets smaller than $3.2~\rearth$ with mass determinations better than 30\% precision. The gray region to the lower right indicate the region where planets would have an iron content exceeding the maximum value predicted from models of collisional stripping \citep{marcus_minimum_2010}. The solid curves are theoretical models for planet with a composition consisting of 100\% water (blue), 25\% silicate and 75\% water (turquoise), 50\% silicate and 50\% water (magenta), 100\% silicate (green), 70\% silicate and 30\% iron consistent with an Earth like composition (light blue), 50\% silicate and 50\% iron (brown) and 100\% iron (red) \citep{zeng_detailed_2013}. The blue points indicate planets with masses measured using the HARPS-N spectrograph, purple points are from other sources and the orange points are the \kep\ system with masses determined in this paper.}
	\end{center}
	\label{fig:mrdiagram}
\end{figure*}

The resulting planetary parameters for the system are detailed in Table \ref{tab:planetparameters} and the masses and radii of the \kep\ planets are shown in a mass-radius diagram in Figure \ref{fig:mrdiagram}. We find the masses of the planets to be $M_{p,b} = \KBmp~\mearth~(\KBmpprecision)$, $M_{p,c} = \KCmp~\mearth~(\KCmpprecision)$, $M_{p,d} = \KDmp~\mearth~(\KDmpprecision)$ and $M_{p,g} = \KGmp~\mearth~(\KGmpprecision)$ (values in parentheses indicate the precision). Both \kep d and \kep c have densities consistent with a  composition comprising a core with an extended H/He atmosphere or a significant fraction of volatiles. However, \kep b has a density consistent with a bare rocky terrestrial composition without volatiles or a H/He atmosphere, despite its relatively large size of $\sim 1.9~\rearth$. The non-transiting planet, \kep g, has a minimum mass similar to Neptune.

To test whether the system configuration of the fits had any influence on the measured masses, we ran the RV MCMC analysis with the following configurations: a 3 planet model imposing circular orbits, a 3 planet model with uniform priors on the eccentricities, a 4 planet model imposing circular orbits, a 4 planet model with uniform priors on the eccentricities and finally a 4 planet model with priors on the eccentricities imposed by the stability analysis in Section \ref{sec:stability} (the adopted analysis). The resulting masses of all the planets in all these various configurations are compatible within their one-sigma uncertainties. The system parameters, including the planet masses, are also compatible within the one-sigma uncertainties if we include the 21 rejected RV gathered under poor observing conditions with internal uncertainties greater than $5~\ms$.

In order to evaluate whether a 3 or 4 planet model is preferred, we fixed the jitter terms to the values reported in the next paragraph and computed the Bayesian Information Criterion for the 3 and 4 planet models. We find a $\Delta \textrm{BIC} = 7.0$, indicating that including the non-transiting planet is the preferred model. We chose to adopt the 4 planet model with priors on the eccentricities from the stability analysis because of the strong evidence for the non-transiting planet presented in Section \ref{sec:nontransit} and the stability arguments in Section \ref{sec:stability}. 

The jitter terms for the two instruments are similar with a HARPS-N jitter term of $\kjitterHARPS~\ms$ and a HIRES jitter term of $\kjitterHIRES~\ms$. The average internal errors are slightly lower for HIRES ($\krverrintHIRES~\ms$) compared to HARPS-N ($\krverrintHARPS~\ms$), which is similarly seen for in the average uncertainties including jitter of ($\krverrtotHIRES~\ms$) for HIRES and ($\krverrtotHARPS~\ms$) for HARPS-N. We attribute the lower uncertainties of the HIRES RVs to the significantly higher SNR of these spectra (over 400\% higher SNR than the HARPS-N spectra).

\begin{deluxetable}{ccc}
	
	\tablecolumns{3}
	\tiny
	\tablecaption{Transit parameters for planets with no mass measurements \label{tab:photoparameters}}
	\tablehead{
		\colhead{Parameter} &
		\colhead{Kepler-20e} &
		\colhead{Kepler-20f}
	}
	\startdata
	$P$ (days) & $6.09852281_{-0.00001351}^{+0.00000608}$ & $19.57758478_{-0.00012256}^{+0.00009037}$ \\
	$T_c$ (BJD - 2,454,000) & $968.931544_{-0.000732}^{+0.002223}$ & $968.207060_{-0.004332}^{+0.006076}$ \\
	$R_p/R_\star$ & $0.00822_{-0.00021}^{+0.00020}$ & $0.00952_{-0.00083}^{+0.00044}$ \\
	$a/R_\star$ &  $14.26_{-0.32}^{+0.42}$ & $31.15_{-0.79}^{+0.81}$ \\
	$i$ (deg) & $87.632_{-0.128}^{+1.085}$ & $88.788_{-0.072}^{+0.426}$ \\	
	$q_1$ & $0.045 \pm 0.188$ & $0.995_{-0.466}^{+0.068}$ \\	
	$q_2$ & $0.015 \pm 0.457$ & $0.765_{-0.444}^{+0.110}$ \\	
	$R_p (R_\oplus)$ & $0.865_{-0.028}^{+0.026}$ & $1.003_{-0.089}^{+0.050}$ \\
	$R_p$ error  (\%) & $3.1\%$ & $6.9\%$ \\
	Semi-major axis $a$ (AU) & $0.0639_{-0.0014}^{+0.0019}$ & $0.1396_{-0.0035}^{+0.0036}$ \\	
	\enddata
	\tablecomments{The values reported are the modes of the posterior distribution and the errors encompass the 68\% of data points closest to the mode values. Planet radius estimates assume a stellar radius of $R_\star = 0.9639 R_\odot \pm 0.0177$.}
	
\end{deluxetable}

\begin{deluxetable*}{lcccc}
	\tabletypesize{\scriptsize}
	\tablecaption{Planet parameters for planets with mass determinations
		\label{tab:planetparameters}}
	\tablehead{
		\colhead{Parameter}				& 
		\colhead{\kep b} 				&
		\colhead{\kep c}				&
		\colhead{\kep g}				&
		\colhead{\kep d}				
	}
	\startdata
	Orbital period $P$ (days)  & \KBperiod & \KCperiod & \KGperiod & \KDperiod \\
	$T_C$ (BJD - 2,454,000) & \KBtc & \KCtc & \KGtc & \KDtc\\
	$R_P/\rstar$ & \KBrprstar & \KCrprstar & \KGrprstar & \KDrprstar \\
	$a/\rstar$ & \KBarstar & \KCarstar & \KGarstar & \KDarstar \\
	$q_1$ & \KBqone & \KCqone & \KGqone & \KDqone \\
	$q_2$ & \KBqtwo & \KCqtwo & \KGqtwo & \KDqtwo \\
	Oribtal inclination $i$ (deg) & \KBinc & \KCinc & \KGinc & \KDinc \\
	Orbital eccentricity $e$ & \KBecc & \KCecc & \KGecc & \KDecc \\
	$\sqrt{e}\sin{\omega}$ & \KBsqrtesinw & \KCsqrtesinw & \KGsqrtesinw & \KDsqrtesinw \\
	$\sqrt{e}\cos{\omega}$ & \KBsqrtecosw & \KCsqrtecosw & \KGsqrtecosw & \KDsqrtecosw \\
	Orbital semi-amplitude $K$ (\ms) & \KBK & \KCK & \KGK & \KDK  \\
	Mass \mpl (\mearth) & \KBmp & \KCmp & \KGmp$^a$ & \KDmp \\
	Mass \mpl~error (\%) & \KBmpprecision & \KCmpprecision & \KGmpprecision & \KDmpprecision \\
	Radius \rpl (\rearth) & \KBrp & \KCrp & \KGrp & \KDrp \\
	Radius \rpl~error (\%) & \KBrperr & \KCrperr & \KGrperr & \KDrperr \\
	Planet density \rhopl ($\mathrm{g~cm^{-3}}$) & \KBrhopl & \KCrhopl & \KGrhopl & \KDrhopl \\
	Orbital semi-major axis $a$ (AU) & \KBa & \KCa & \KGa & \KDa \\
	Equilibrium temperature $T_{eq}$ (K) $^b$& \KBteq & \KCteq & \KGteq & \KDteq \\	
	\enddata
	\tablenotetext{a}{
		Minimum mass ($\msini$), since the inclination of the orbit of \kep g is not known.
	}
	\tablenotetext{b}{
		Assuming a Bond albedo of 0.3.
	}
\end{deluxetable*}

\begin{figure*}
	\gridline{\fig{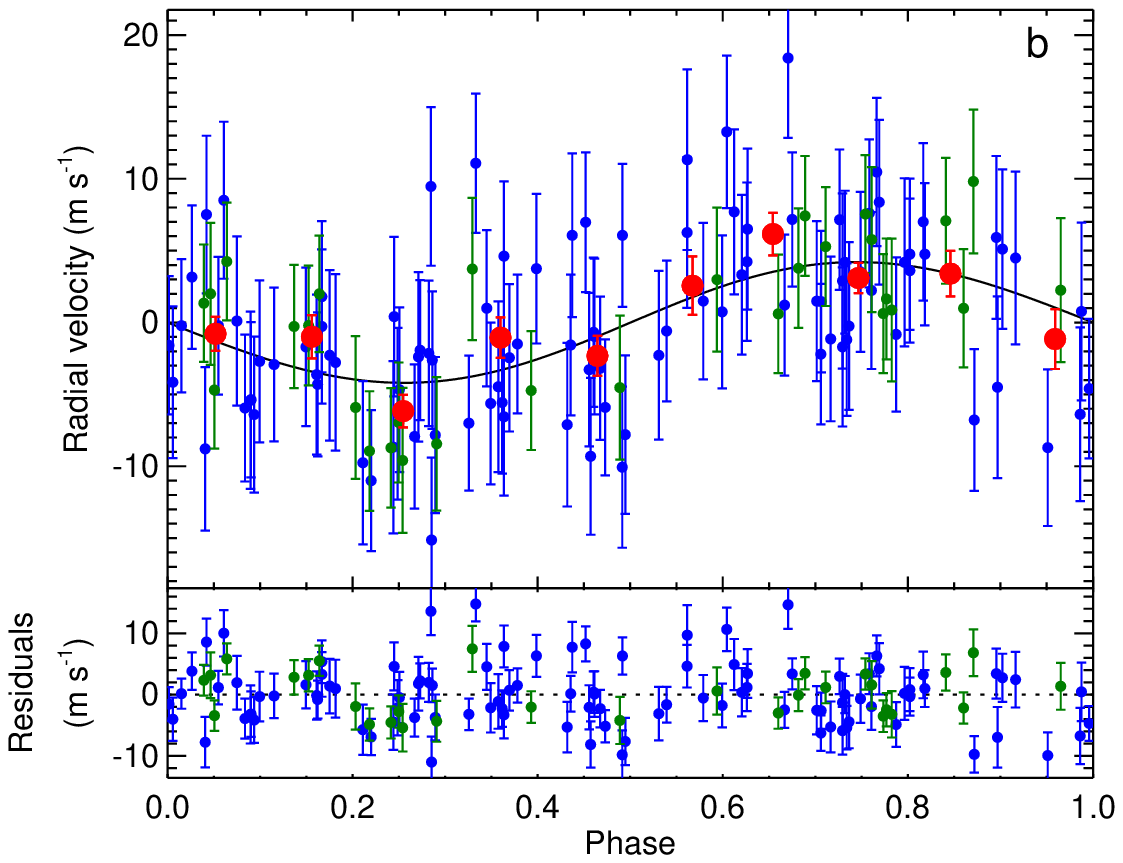}{0.5\textwidth}{}
		\fig{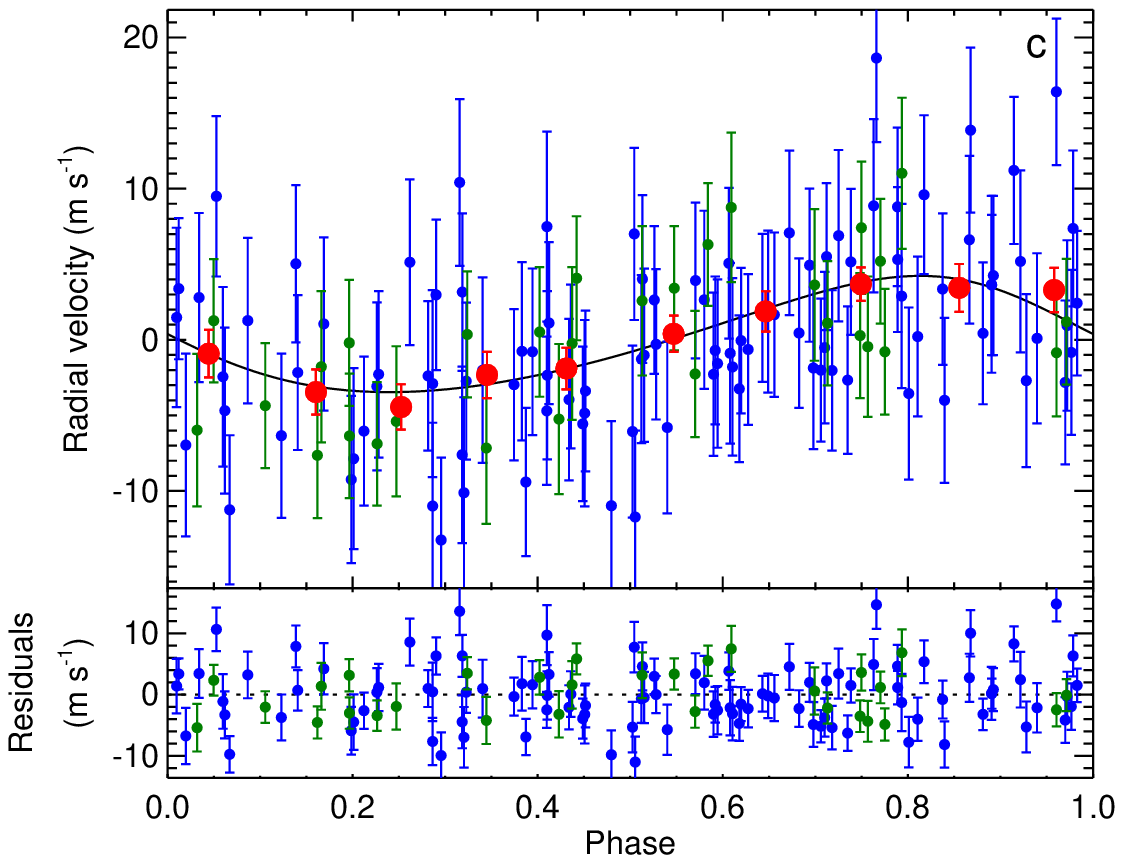}{0.5\textwidth}{}}
	\gridline{\fig{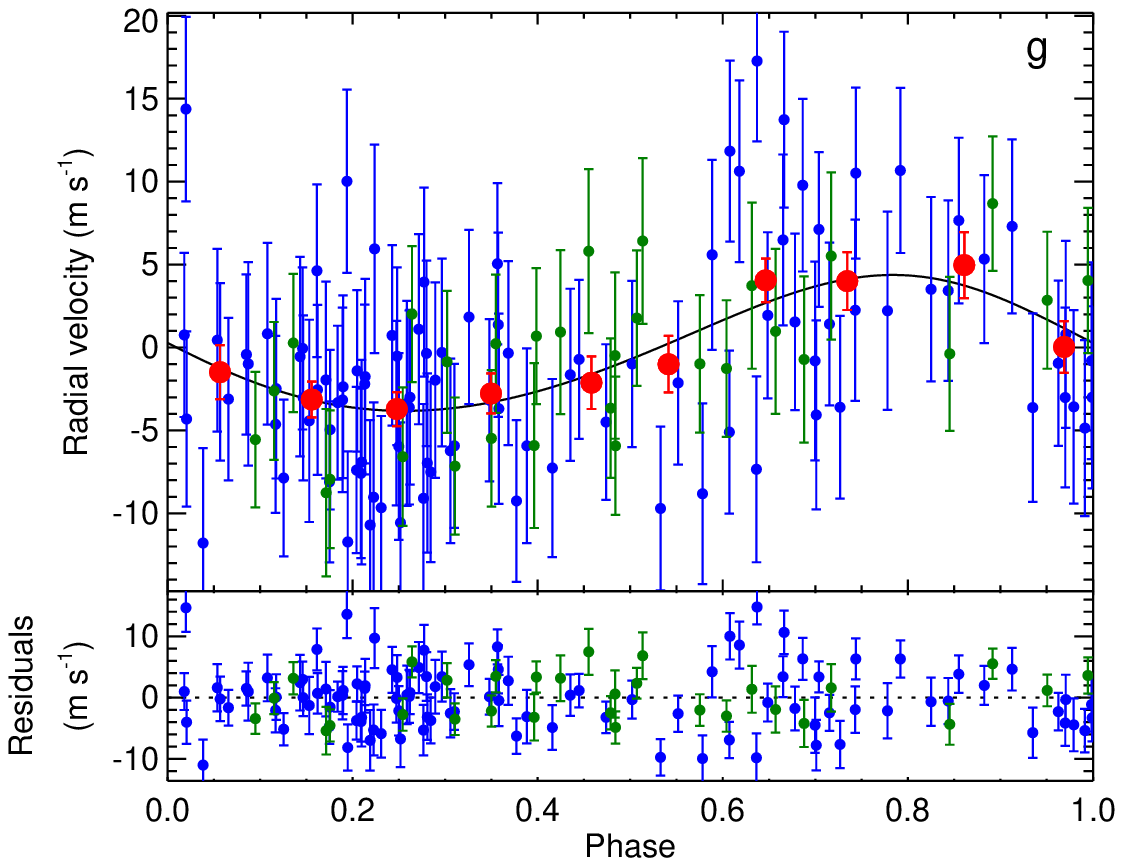}{0.5\textwidth}{}
		\fig{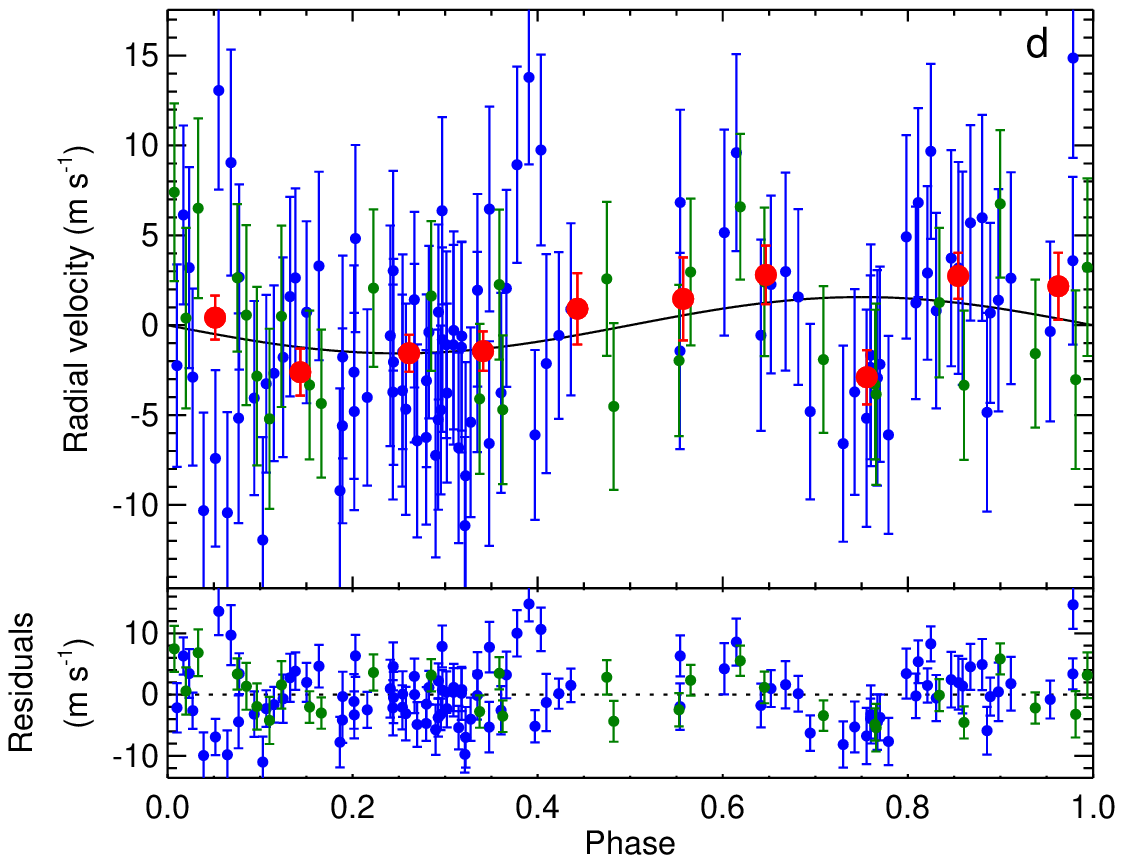}{0.5\textwidth}{}}
	\caption{The best--fit model (black solid curve) to the HARPS-N (blue points) and HIRES (green points) radial velocities, when considering a model composed of the three largest transiting planets and the non-transiting planet in the Kepler-20 system. Each top panel show the phase-folder RVs after removing the orbital motion of the remaining planets. The red points show the weighted mean of the RVs binned to equal interval in phase. Each bottom panel shown the RV residuals after removing the full orbital fit. The four plots show the orbital fit for \kep b (top-left), \kep c (top-right), \kep g (bottom-left), \kep d (bottom-right).}
	\label{fig:orbitfit}
\end{figure*}

\section{The non-transiting planet \kep g}
\label{sec:nontransit}

\begin{figure}[htbp]
	\plotone{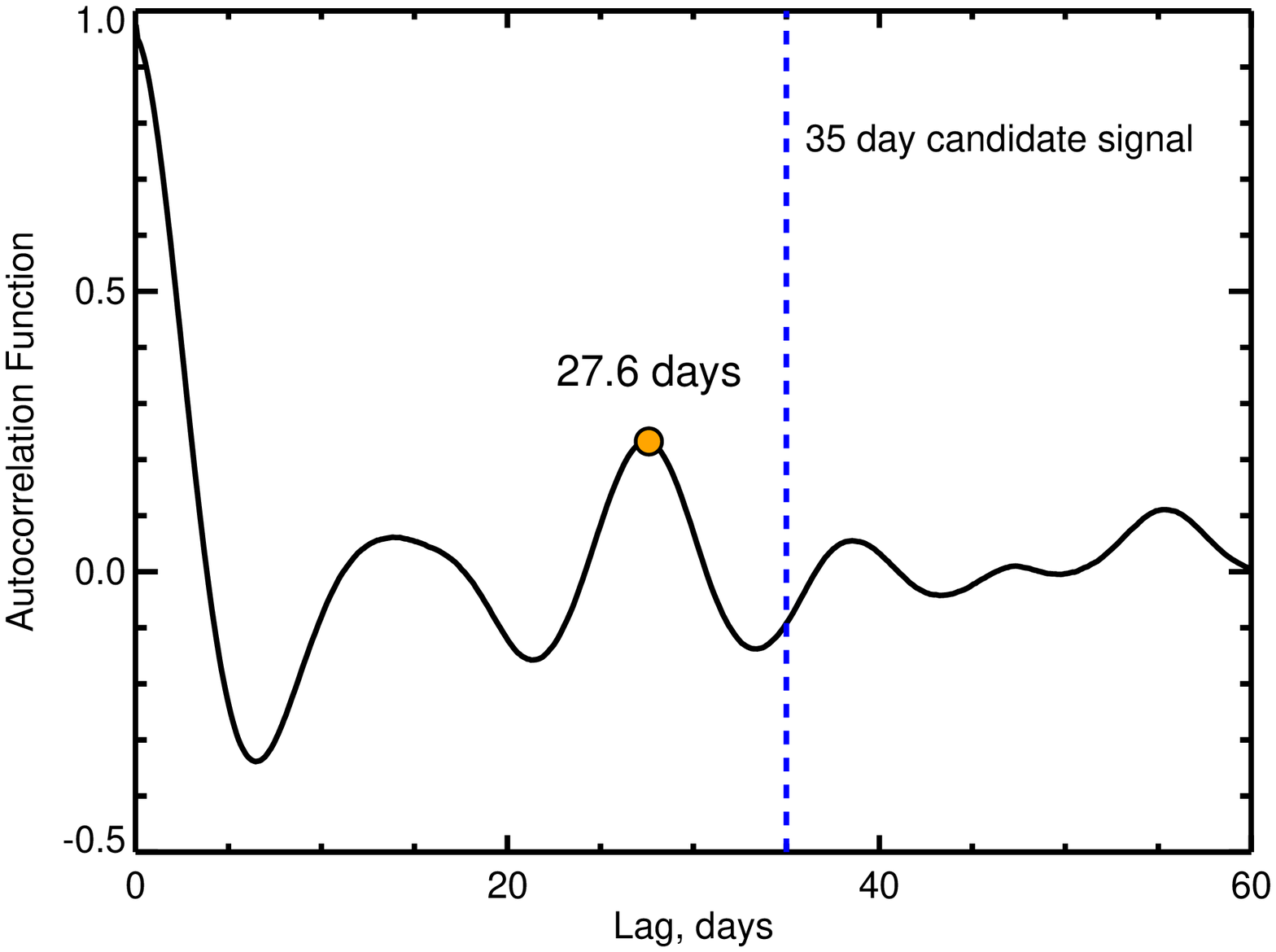}
	\caption{Autocorrelation function of the Kepler-20 light curve. There is a peak in the autocorrelation function at 27.6 days, which we interpret as the stellar rotation period. This is significantly different from the 35 day period of the RV variations which we interpret as a new non-transiting planet (shown as a blue dashed line).}
	\label{fig:acf}
\end{figure}

In this section we argue that the planetary interpretation of the 35 day signal is more likely than a stellar activity interpretation. We measured the rotation period of \kep\ from the Kepler PDC-SAP light curves using an autocorrelation function (ACF) analysis \citep[e.g.][]{mcquillan_stellar_2013}. We calculated the ACF for the full dataset (quarters 1-17) and found a strong peak at a period of 27.6 days (see Figure \ref{fig:acf}). The signal near 27.6 days is persistent -- we divided the dataset in two, repeated the autocorrelation analysis on each half of the light curve, and found a peak at 27.0 days in the first half and a peak at 27.7 days in the second half. We interpret this 27.6 day signal as being caused by active regions on the stellar surface, and measure the rotation period of \kep\ to be $27.6 \pm 0.5~\days$. Furthermore, a periodogram of the FWHM measurements of the CCF function shows significant power at 27 days, indicating a stellar rotational period similar to that found from the photometry.

We believe it is unlikely that the period of the additional RV signal at 35 days is consistent with the stellar rotation period. The HIRES radial velocity data were taken concurrently with Kepler observations (during the first half of the full Kepler dataset). During this time, active regions on the stellar surface gave rise to a signal with a period of 27.0 days in the flux time series, while the additional RV signal (which was detected in the HIRES data alone) has a period of 35 days. These periods differ by 20\%, and the ACF shows only a weak (negative) correlation between flux measurements taken 35 days apart.

However, in order to provide further evidence for the planetary nature of the signal, we perform further tests to show that it is not related to stellar activity. To do so,  we removed the signature of Kepler-20b, c, and d from the RVs, using our best fit, leaving only the 35-day signal. We then split the RV measurements in two equal chunks and estimated the generalized Lomb-Scargle (GLS) periodogram \citep[][]{zechmeister_generalised_2009} for each of these chunks. Note that the power in a GLS periodogram depends on the sampling and the total time-span of the data. Therefore, to be able to compare the signals found in the two chunks, we needed to keep the same sampling and timespan. For each chunk, we therefore kept the RV measurements outside of the chunk, but fixed their value to zero, with error bars of 100 \ms. A similar analysis was performed for $\alpha$ Cen Bb \citep{dumusque_earth-mass_2012} and Kepler-10 \citep{dumusque_kepler-10_2014}. In Figure \ref{fig:xdum_fig0}, we show the periodogram for each chunk of data (blue and red) and the GLS periodogram for the entire data set (gray). The signal at 35 days is seen in the first and second halves of the data, which is expected for a signal induced by a planet or by stellar activity if 35 days is the stellar rotation period. However, when looking at the phase of the signal, illustrated in Figure \ref{fig:xdum_fig0} by small arrows above the 35-day peak, it is clear that the signal has the same phase in each chunk. Furthermore, this phase is consistent with the phase derived when analyzing the entire data set. The signal thus retains the same phase from season to season, which is a strong argument in favor of the planetary origin of the signal. A signal due to stellar activity would change its phase because of the evolution of active region configuration on the stellar surface. 

\begin{figure}
	\begin{center}
		\includegraphics[width=8cm]{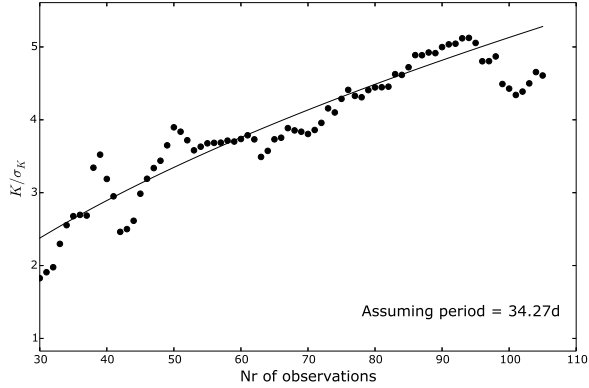}
		\caption{SNR of the semi-amplitude for a signal with period $P = 34.27~\days$ (assuming a circular orbit) versus number of data points. The solid line represents a square root function with $y = 0.61\sqrt{x} - 0.96$ (best fit to the data).}
		\label{fig:periodi_growth}
	\end{center}
\end{figure}

To further assess the strength of the detection of the non-transiting planet, we calculated the SNR of the semi-amplitude of the signal for subsamples of the data. We first performed the Bayesian General Lomb-Scargle periodogram \citep[BGLS; ][]{mortier_bgls:_2015} on the full dataset. The strongest period in the BGLS, which assumes a circular orbit, is $34.27~\days$. We then created subsamples of the data, starting with the first 30 data points and always adding one more data point. For each subsample, and fixing the period to $34.27~\days$, we found the best jitter term, by calculating the maximum log likelihood of the model for a set of jitter values (spaced with $0.1~\ms$). Using this jitter term, we derived optimal scaling estimates for the semi-amplitude $K$ and its variance $\sigma^2_K$, ultimately giving the SNR of the signal over time: $K/\sigma_K$. From Figure \ref{fig:periodi_growth}, it can be seen that the SNR roughly scales as a square root, as to be expected for a coherent signal in the data.

In summary, the tests described in this section strongly indicate that the signal at $\sim35~\days$ is of planetary nature and we therefore conclude that a non-transiting planet is present in the system with a period of $\KGperiod~\days$.

\section{Stability analysis}
\label{sec:stability}

To investigate the stability of the Kepler-20 system, we carried out pure N-body simulations using mercury6 \citep{chambers_hybrid_1999}.  We also carried out some tests using a 4th order Hermite integrator \citep{makino_hermite_1992}, but the results from this were consistent with those from mercury6, and so we report here only the results from the mercury6 simulations. We considered all 6 bodies, Kepler-20b, c, d, e, f, and g, orbiting a central star with mass $\mstar = \kepmstar~\msun$.  The masses of Kepler-20b, c, d and g are taken from Table~\ref{tab:planetparameters}, while Kepler-20e is assumed to have a mass of $\mpl = 0.65~\mearth$ and Kepler-20f is assumed to have a mass of $\mpl = 1.0~\mearth$.

The initial semi-major axes are the same as in Table~\ref{tab:planetparameters}, and the system is assumed to be co-planar. The separation between adjacent planets, normalized to the mutual Hill radius, lies between 13 and 20, suggesting that the system lies outside the empirical stability limit of $\sim 10$ \citep{chambers_stability_1996,yoshinaga_stability_1999}, and should therefore be long-term stable if the eccentricities are sufficiently small.  In our first test, we set the initial eccentricity for Kepler-20b to $e_b = 0.03$ and set all the other eccentricities to $e = 0$. With these initial conditions, the system is indeed stable for $t > 10^7$ years ($> 5 \times 10^8$ orbits of Kepler-20d).

For Kepler-20c, the system is stable for eccentricities of $e_c \le 0.17$, for Kepler-20g
the system is stable for eccentricities of $e_g \le 0.16$, while for Kepler-20d the system is stable for at eccentricities of $e_d \le 0.28$. Essentially, Kepler-20c and Kepler-20g appear to be constrained to have eccentricities well below $e = 0.2$, while Kepler-20d may have an eccentricity as high as $e \sim 0.3$. 

We also considered scenarios where we varied the eccentricities of more than one of the planets. We, again, ran the simulations for $10^7$ years ($5 \times 10^8$ orbits of Kepler-20d). If Kepler-20c has an eccentricity of $e_c = 0.1$, then the system is only stable if Kepler-20g has an eccentricity of $e_g \le 0.11$. If Kepler-20g has an eccentricity of $e_g = 0.1$ then the system is only stable if Kepler-20c has an eccentricity of $e_c \le 0.11$. If both Kepler-20c and Kepler-20g have eccentricities of $e_{c,g} = 0.1$, then the system is stable if Kepler-20d has an eccentricity of $e_d \le 0.22$. As expected, if the eccentricities are non-zero, then the range of eccentricities for which the system is stable is reduced. 

We should acknowledge that we haven't run these simulations for longer than $10^7$ years. Hence, the system could still be unstable for the eccentricity values we present here and, consequently, these should be regarded as upper limits. However, they do indicate that the eccentricities of Kepler-20c and g are likely to be constrained to be less than $e \sim 0.2$, while the eccentricity of Kepler-20d could be higher, but only if the others have relatively low eccentricities ($e \sim 0.1$ or less).

Overall, the stability analysis appears to suggest that Kepler-20 is consistent with being a dynamically cold system in which the eccentricities and inclinations are small \citep{dawson_correlations_2016}, and that probably formed in the transition phase when the disc had a high solid surface density, but a low to moderate gas surface density \citep{lee_breeding_2016}.

\section{Summary and Discussion}
\label{sec:discussion}

We have revisited the mass determination of the planets in the \kep\ system. With a significantly increased number of RV measurements, an updated photometric analysis of the Kepler light curve, and an asteroseismic analysis of the parameters of the host stars, we present mass measurements of the three larger planets in the \kep\ system as well as the detection of a non-transiting planet. With the refined analyses, we are able to reduce the uncertainty of the mass measurement of \kep b to less than 15\%. \kep b has a mass of $M_{P,b} = \KBmp~\mearth~(\KBmpprecision)$ and a radius of $R_{P,b} = \KBrp~\rearth$ yielding a bulk density of $\rho_b = \KBrhopl~\gcmc$. The precision allows us to conclude that \kep b has a composition consistent with a rocky terrestrial composition despite it large radius ($\sim 1.9~\rearth$). We also report the masses of \kep c and \kep d ($M_{p,c} = \KCmp~\mearth~(\KCmpprecision)$, $M_{p,d} = \KDmp~\mearth~(\KDmpprecision)$) yielding densities indicating the presence of volatiles and/or a H/He atmosphere.

Furthermore, we report the detection of a non-transiting planet in the system (\kep g) with a minimum mass similar to that of Neptune ($M_{p,g} = \KGmp~\mearth~(\KGmpprecision)$). The orbital configuration of the \kep\ system has a rather larger gap between \kep f and \kep d in which \kep g is situated with an orbital period of $\sim 35~\days$. In Section \ref{sec:nontransit} we find strong evidence for the signal at $\sim 35~\days$ to be of planetary nature and not the result of stellar activity caused by spot modulation at the stellar rotational period.

\kep b is thus the first definite example of a rocky exoplanet with a mass above $5~\mearth$ and a radius larger than $1.6~\rearth$. This is seemingly at odds with the prediction that most $1.6~\rearth$ planets are not rocky and larger planets are even more likely to exhibit lower bulk densities \citep{rogers_most_2015,weiss_mass-radius_2014}. However, it is worth noting that all seven planets in Figure \ref{fig:mrdiagram} with densities consistent with a rocky composition (Kepler-10b, Kepler-20b, Kepler-36b, Kepler-78b, Kepler-93b, HD 219134b and CoRoT-7b) are highly irradiated with a bolometric flux larger than $F_v > 2 \cdot 10^{5}~\mathrm{J~s^{-1}~m^2}$ ($> 146$ times the flux received by the Earth). It is thus possible, and perhaps even likely, that we are measuring the masses of small planets in a particular part of parameter space that are hot and highly irradiated and could have lost their primordial gaseous envelope (if such envelopes were present) due to photo evaporation. If this is the case, we are thus measuring the masses of the bare cores of these planets. 

Indeed, given the close-in orbit of Kepler-20b, and the likely low masses of Kepler-20e \& f, we conclude that all three of these now likely rocky planets, could have formed with significant gaseous envelopes which were subsequently lost to atmospheric photo-evaporation \citep[e.g.][]{lopez_how_2012} or other processes like impact erosion \citep[e.g.][]{inamdar_formation_2015}. Using the coupled thermal and photo-evaporative planetary evolution model of \citet{lopez_how_2012}, we find that the all the planets in the Kepler-20 system are consistent with a scenario in which each of the planets accreted  a significant H/He envelope composing $~2-5\%$ of its total mass, but planets b, e, \& f lost their in envelopes due to subsequent evolution. This is similar to the results found for a handful of other systems like Kepler-36 \citep[e.g.][]{lopez_role_2013}, however, this type of evolutionary analysis is only possible for transiting systems with precise mass measurements.

It is important to significantly increase the number of precise mass measurements of transiting planets in order to populate the mass--radius diagram with planets of different sizes and masses at various orbital distances, including longer periods, that orbit a diverse sample of stellar types in order to fully appreciate the nature of the small exoplanets and their compositions.


\begin{acknowledgements}
The HARPS-N project was funded by the Prodex Program of the Swiss Space Office (SSO), the Harvard- University Origin of Life Initiative (HUOLI), the Scottish Universities Physics Alliance (SUPA), the University of Geneva, the Smithsonian Astrophysical Observatory (SAO), and the Italian National Astrophysical Institute (INAF), University of St. Andrews, Queen's University Belfast and University of Edinburgh. The research leading to these results has received funding from the European Union Seventh Framework Programme (FP7/2007-2013) under Grant Agreement No. 313014 (ETAEARTH).

This work was performed in part under contract with the California Institute of Technology/Jet Propulsion Laboratory funded by NASA through the Sagan Fellowship Program executed by the NASA Exoplanet Science Institute.

A.V. is supported by the NSF Graduate Research Fellowship, Grant No. DGE 1144152.

This publication was made possible by a grant from the John Templeton Foundation. The opinions expressed in this publication are those of the authors and do not necessarily reflect the views of the John Templeton Foundation. This material is based upon work supported by NASA under grant No. NNX15AC90G issued through the Exoplanets Research Program.

Funding for the Stellar Astrophysics Centre is provided by The Danish National Research Foundation (Grant agreement no.: DNRF106). The research is supported by the ASTERISK project (ASTERoseismic Investigations with SONG and Kepler) funded by the European Research Council (Grant agreement no.: 267864). M.S.L is supported by The Danish Council for Independent Research's Sapere Aude program (Grant agreement no.: DFF -- 5051-00130).

PF acknowledges support by Funda\c{c}\~ao para a Ci\^encia e a Tecnologia (FCT) through Investigador FCT contract of reference IF/01037/2013 and POPH/FSE (EC) by FEDER funding through the program ``Programa Operacional de Factores de Competitividade - COMPETE'', and further support in the form of an exploratory project of reference IF/01037/2013CP1191/CT0001.

The research leading to these results also received funding from the European Union Seventh Framework Programme (FP7/2007- 2013) under grant agreement number 313014 (ETAEARTH).

XD is grateful to the Society in Science$-$Branco Weiss Fellowship for its financial support.
\end{acknowledgements}


\bibliography{lblib}

\end{document}